\documentclass[twocolumn,showpacs,amsmath,amssymb]{revtex4}
\usepackage[dvips]{graphicx}
\usepackage{graphics}
\usepackage{dcolumn}
\usepackage{bm}
\usepackage{bbm}

\begin{document}
\title{Bose-Einstein  condensate of two-dimensional excitons in a ring:
Necklace-like modulation of order parameter }
\draft
\author{S.-R. Eric Yang}
\email{eyang@venus.korea.ac.kr}
\author{Q-Han\ Park}
\email{qpark@korea.ac.kr}
\affiliation{Department of Physics, Korea University, Seoul 136-701, Korea} 
\author{J. Yeo}
\email{jhyeo@konkuk.ac.kr}
\affiliation{Department of Physics, Konkuk University, Seoul 143-701, Korea}


\begin{abstract}
We have studied theoretically the
Bose-Einstein condensation (BEC)
of two-dimensional  
excitons in a ring with small random width variation.
We derive a nonlinear Gross-Pitaevkii equation (GPE)
for such a condensate.  Our numerical solution of the ground state of the GPE
displays a necklace-like structure in the presence of  small random variation of ring width.
This is a consequence of the interplay between random potential   
and the nonlinear repulsive term of  the
GPE.  
Our result suggests that the formation of ring and necklace-like 
structures observed recently in the photoluminescence of quantum wells may be
stable even in the BEC phase.
\end{abstract}

\thispagestyle{empty}
\maketitle

Recent advances in the Bose-Einstein
condensation (BEC) \cite{smith,griffin,mosk} of atoms were made possible by exploring the confinement 
of atomic gases within potential traps \cite{wieman}.
In semiconductor physics, the possibility of exciton BEC has been explored for a long time \cite{griffin,mosk}.
In the dilute limit $na_B^D<<1$ ($a_B$ is the  exciton Bohr radius, $D$ the dimensionality, and $n$ 
is the exciton density) excitons are expected to undergo the BEC \cite{keldysh}.
Recently, the possibility of BEC of excitons in potential traps of semiconductor  quantum 
wells has been explored  \cite{zhu,butov1,butov2,butov3,snoke}. 
The quantum degeneracy temperature is expected to be about 1K, much higher than that of atoms due to
the small mass of the exciton.
Butov et al.~\cite{butov2,butov3}, Snoke et al.~\cite{snoke}, and Rapaport et al.~\cite{rapa} have observed 
a circular
ring in exciton photoluminescence (PL) from these quantum wells, where the radius of the ring
(typically $10\sim 100 \mu{\rm m}$)
depends on the laser excitation intensity.
A mechanism for the  formation of the ring is proposed that involves 
carrier imbalance, transport, and recombination \cite{butov3,rapa}.
In addition, Butov et al.~\cite{butov2} have observed 
that, as temperature decreases, the exciton PL fragments suddenly into a macroscopically ordered state of
nearly necklace-like structure:
they  observed 
that minimum and maximum positions of PL intensity  vary nearly periodically  along the ordered ring,
although the intensity of minimum and maximum values are non-periodic.  The distance between 
peak positions deviate from the mean period roughly about 20 percent. (For a typical large ring
observed in Ref.~\cite{butov2},
the distances between peak positions vary  roughly between $14-22\mu{\rm m}$.) These random
variations are not understood yet.  

In this Letter, we derive a nonlinear Gross-Pitaevkii equation (GPE) of exciton BEC  in the presence 
of  small random variation of ring width, and show that the ground state 
displays a necklace-like structure.  
Following the experimental observation of Butov et al.~\cite{butov2} and Snoke et al.~\cite{snoke}
we assume that electrons and holes are confined to move in a ring.
The observed  ring radius and width vary locally \cite{com1}.
For simplicity we will model this randomness  through  a  random  variation of ring width
as excitons move along the circumference of the ring with a constant radius.
This implies that excitons experience a random potential 
as they move along the circumference (smaller width
gives larger in-plane exciton potential energy).
In this paper we assume that excitons form a Bose-Einstein condensate at temperatures  below
the quantum degeneracy temperature and explore the nature of such a condensate
in the  presence of a  random  potential.
We employ a mean field approach \cite{keldysh,zhu}, which is exact both 
in the low and high density limits.   We  derive in the low density limit a 
nonlinear Gross-Pitaevskii equation (GPE) for excitons  
and solve  it numerically for the ground state using the imaginary time evolution method. 
We find that 
the order parameter of exciton BEC displays a necklace-like structure. 
This is a consequence of a nontrivial interplay between a
repulsive nonlinear term of GPE and  random  potential.
Our result suggests that 
the formation of   a necklace-like structure
observed recently in the PL of 
quantum wells may be
stable even in the  BEC phase at lower temperatures.

We derive an effective theory of the BEC
of two-dimensional excitons in a potential trap, which will lead to
the GPE.
We perform a model calculation in which
the electron-hole interaction to be $-g\delta({\bf x}-{\bf x}'),~g>0$. 
We only consider optically active s-wave excitons and ignore heavy and light hole coupling 
in the valence band \cite{yang}. 
We replace the quartic electron-hole interaction 
in the Hamiltonian involving the electron and
hole operators, $\psi_e({\bf x})$ and $\psi_h({\bf x})$, with
those containing the pairing averages, $g\langle \psi_h ({\bf x}) \psi_e ({\bf x})\rangle$,
and $g\langle\psi^\dagger_e ({\bf x}) \psi^\dagger_h ({\bf x}) \rangle$ denoted respectively
by the gap functions, $\Delta ({\bf x})$ and $\Delta^*({\bf x})$.

\newpage

The system is described by the finite-temperature Green's functions defined by
$G_e ({\bf x}\tau,{\bf x}^\prime\tau^\prime)=-\langle T_\tau\psi_e({\bf x},\tau)
\psi^\dagger_e ({\bf x}^\prime,\tau^\prime) \rangle$,
$G_h ({\bf x}\tau,{\bf x}^\prime\tau^\prime)=-\langle T_\tau\psi_h({\bf x},\tau)
\psi^\dagger_h ({\bf x}^\prime,\tau^\prime) \rangle$,
$F({\bf x}\tau,{\bf x}^\prime\tau^\prime)=-\langle T_\tau\psi_e({\bf x},\tau)
\psi_h ({\bf x}^\prime,\tau^\prime) \rangle$, and 
$F^\dagger({\bf x}\tau,{\bf x}^\prime\tau^\prime)=-\langle T_\tau\psi^\dagger_h({\bf x},\tau)
\psi^\dagger_e ({\bf x}^\prime,\tau^\prime) \rangle$,
where $T_\tau$ means the time ordering in the imaginary time $\tau$.
These Green's functions satisfy
\begin{widetext}
\begin{eqnarray}
\left[
\begin{array}{cc}
-\hbar\frac{\partial}{\partial\tau}-\frac{1}{2m_e}(-i\hbar\nabla+\frac{e}{c}{\bf A})^2+\mu_e-V_e({\bf x})
& \Delta({\bf x})\\
\Delta({\bf x})^*  &  -\hbar\frac{\partial}{\partial\tau}+\frac{1}{2m_h}(i\hbar\nabla-\frac{e}{c}{\bf A})^2
-\mu_h+V_h({\bf x})   \\
\end{array}\right]\nonumber\\
\times\left[\begin{array}{cc}
G_e({\bf x}\tau ,{\bf x}'\tau') & F({\bf x}\tau, {\bf x}'\tau')\\
 F^\dagger ({\bf x}\tau, {\bf x}'\tau')  & -G_h( {\bf x}'\tau',{\bf x}\tau )
\end{array}\right]=
\hbar\delta({\bf x}-{\bf x}')\delta(\tau-\tau') \mathbbm{1},
\end{eqnarray}
\end{widetext}
where
the gap function $\Delta({\bf x})=gF({\bf x}\tau^+, {\bf x}\tau)$ is obtained from the Green's function
self-consistently.
Here $V_{e,h}({\bf x})$ is the external potential for the electrons and holes set by the potential trap
as well as by the random disorder.
In this model, we also consider a simple situation where
the effect of the electron-electron and the hole-hole interactions
can be included in $V_{e,h}({\bf x})$.
Note the presence of the magnetic field represented by
the vector potential ${\bf A}$.

The coupled differential equations can be  written as integral equations using the noninteracting
Green's functions, $\tilde{G}^0_{e,h}$ that are obtained for the case where $\Delta=0$.
Iterating the integral equations and keeping up to, say, the third order in $\Delta$, we find
\begin{eqnarray}
&&g^{-1}\Delta^*({\bf x})\approx\int d^2{\bf y}\;Q({\bf x},{\bf y})
\Delta^*({\bf y}) \label{gpdelta} \\
&&+\int\prod^3_{i=1}d^2 {\bf y}_i\;
R({\bf x},{\bf y}_1,{\bf y}_2,{\bf y}_3)\Delta^*({\bf y}_1) 
\Delta({\bf y}_2) \Delta^*({\bf y}_3), 
\nonumber
\end{eqnarray}
where
$Q({\bf x},{\bf y})=(\beta \hbar^2)^{-1}\sum_n \tilde{G}^0_h({\bf y},{\bf x},-\omega_n)
\tilde{G}^0_e({\bf y},{\bf x},\omega_n)$ and 
\begin{eqnarray*}
&&R({\bf x},{\bf y}_1,{\bf y}_2,{\bf y}_3)=
-(\beta \hbar^4)^{-1}\sum_n \tilde{G}^0_h({\bf y}_1,{\bf x},-\omega_n)  \\
&&\times\tilde{G}^0_e({\bf y}_1,{\bf y}_2,\omega_n)\tilde{G}^0_h({\bf y}_3,{\bf y}_2,-\omega_n)
G^0_e({\bf y}_3,{\bf x},\omega_n)
\end{eqnarray*}
with the fermionic Matsubara frequency $\omega_n$. The higher order terms in the expansion
are given in a similar way. The magnitude of the higher order terms will be estimated below. 
It turns out that it is sufficient to keep only up to
$O(\Delta^3)$ in the present problem.
Unlike in the derivation of the Ginzburg-Landau equations,
our GPE is valid in the so-called strong coupling regime \cite{drech,haussmann,pieri,pieri2,schakel}, which 
is characterized by
the negative chemical potential $\mu=\mu_e+\mu_h<0$ near $T=0$, or $\beta\mu\to -\infty$.  
We assume that the effect of the magnetic field can be included as a phase factor as
$\tilde{G}^0_{e,h}({\bf x},{\bf y},\omega_n)
=\exp(-i\phi_{e,h}({\bf x},{\bf y}))G_{e,h}^0({\bf x},{\bf y},\omega_n),
$
where $\phi_{e,h}({\bf x},{\bf y})=\exp(\pm\frac{e}{2\hbar c}[{\bf A}({\bf x})+
{\bf A}({\bf y})]\cdot({\bf x}-{\bf y}))$ with the plus (minus) sign
corresponding to the
electrons (holes) and $G_{e,h}^0$ are the
noninteracting Green's function
in the absence of the magnetic field.
This approximation is exactly the eikonal approximation
made by Gorkov in the weak coupling limit \cite{fw}.
This is reasonable in this case also, since ${\bf A}$ can be considered constant over the 
the size of the exciton.
It neglects, however, the terms
of order ${\bf A}^2$ \cite{fw}, thus not suitable for the strong field regime.
Within this approximation, the effect of the magnetic
field disappears in the expressions of
$Q$ and $R$ in the resulting GPE.
Note that, for $V_{e,h}({\bf x})=0$,
$G_{e,h}^0({\bf x},{\bf y},\omega_n)\sim \exp(\sqrt{2m_{e,h}(\mu_{e,h}+i\omega_n})|{\bf x}-{\bf y}|)$,
which decays
within the length scale $\sqrt{\hbar^2/(2m_{e,h}|\mu_{e,h}|)}$.
The presence of an external potential such as $V_{e,h}({\bf x})$ in the strong coupling
limit was studied in Ref.~\cite{pieri}. As long as the external potentials vary slowly over the
Bohr radius and they are much smaller than the binding energy $\epsilon_0$, one can
replace $\mu_{e,h}$ by $\mu_{e,h}-(V_{e,h}({\bf x})+V_{e,h}({\bf y}))/2$ in the Green's functions.

Now, since $Q$ and $R$ are rapidly
decaying
functions of their arguments, we may approximate
the first term on the right hand side of Eq.~(\ref{gpdelta}) as
\[
\Delta^*({\bf x})\int d^2{\bf z} Q({\bf x},{\bf z})+
\frac{1}{4}\Big[\nabla^2 \Delta^*({\bf x})\Big] \int d^2{\bf z}Q({\bf x},{\bf z})({\bf x}-{\bf z})^2,
\]
and the second term as
\[
\Delta^*({\bf x})|\Delta({\bf x})|^2\int\prod^3_{i=1}d^2 {\bf y}_i\;
R({\bf x},{\bf y}_1,{\bf y}_2,{\bf y}_3).
\]
The above integrals are evaluated in the strong coupling limit
at $T=0$ with the assumption
$\mu_{e,h}<0$. 
In our model, excitons move in 
a Mexican hat potential in addition to a random potential represented by $V_D({\bf r})$.
The resulting GPE is conveniently expressed in terms of the pair wavefunction 
$\Phi({\bf x})=\sqrt{\frac{2D}{|\mu|}}\Delta({\bf r})$ where
$D=\frac{m^*}{2\pi\hbar^2}$ is
the two-dimensional
density of state per spin for the  reduced mass $m^*$ given by $1/m^*=1/m_e+1/m_h$.
The evaluation of higher order coefficients in the GPE is straightforward. Here we present the 
GPE up to $O(\Phi^7)$ as follows:
\begin{eqnarray}
\mu_B \Phi({\bf r})&=&-\frac{\hbar^2}{2M}\nabla^2\Phi({\bf r})+
\frac{1}{2}\alpha(r^2-R^2)^2\Phi({\bf r}) \nonumber \\
&&+V_D({\bf r})\Phi({\bf r})
+\frac{1}{2D}|\Phi({\bf r})|^2\Phi({\bf r}) \\
&&-\frac{3}{8|\mu|D^2}
|\Phi({\bf r})|^4\Phi({\bf r})+\frac{1}{12|\mu|^2 D^3}|\Phi({\bf r})|^6\Phi({\bf r}), \nonumber
\end{eqnarray}
where $R$ is the radius of the ring, $M=m_e+m_h$
and $\alpha$ determines the strength of the confinement potential.  
Note that the present GPE is obtained up to $O(\Phi^7)$ in contrast
to the conventional one \cite{smith}, which contains only $O(\Phi^3)$ nonlinear term. 
The $O(\Phi^5)$ term has the opposite sign
to the $O(\Phi^3)$ term and can become important when $\mu$ is small.
This is, however, not the case in the present problem as can be seen below. 
In general,
the magnitude of a higher order nonlinear term may be estimated by multiplying 
the previous lower order term
by $-\frac{1}{D|\mu|}$. The relative importance of each nonlinear term will be
discussed below.
The  condensate
wavefunction is normalized  to  the electron density,  given by $n({\bf x})=|\Phi({\bf x})|^2$.
From the kinetic term of the GPE, we note that
the effective mass of an exciton is
$M=m_e+m_h$, as expected.  Note that the coefficient of the cubic term in $\Phi$
is independent of the binding energy, which is special to two dimensions.
The GPE also shows that,
in the weak magnetic field regime, the exciton does not couple to the vector potential.
Microscopic parameters appear in the energy $\mu_B$, which is evaluated to be
$\mu_B=|\mu|(\ln\frac{\epsilon_D}   {|\mu|}-\frac{1}{gD})
=|\mu|\ln\left(\frac{\epsilon_0}{|\mu|}\right)$,
where 
$\epsilon_0=\epsilon_D e^{-\frac{1}{gD}}$ is the binding energy of the electron-hole pair 
and
$\epsilon_D=\frac{\Lambda^2\hbar^2}{2m^*}$ with the ultraviolet cutoff $\Lambda$
used to regularize the delta-function interaction of the electrons and holes.
This expression for the binding energy can be derived \cite{schakel}
from the Schr\"odinger equation for the particle with reduced mass $m^*$
and energy $-\epsilon_0$
interacting with the delta function potential of strength $-g$. 
The strong coupling limit is defined by the chemical potential $\mu=\mu_e+\mu_h\to -\epsilon_0$.
In the limit $\mu\to-\epsilon_0$,
$\mu_B \simeq
\epsilon_0+\mu$.
This is in contrast with the weak coupling BCS limit where $\mu$ is approximated by the Fermi energy.

We rescale  the wavefunction
$\Phi=A\Psi $ and  the position coordinate 
$x=r/a$ using a length scale $a$.
We define $\hbar\Omega$ 
by the energy scale associated
with the length scale $a$: $a^2=\frac{\hbar}{M\Omega}$.
The energy scale of the confinement potential is $\hbar\omega=\alpha a^4$.
Note that this confinement potential alone does not fix 
the width of the ring since the width depends also on the third order coefficient in $\Phi$ (see below).
Since the number of exciton is $\int|\Phi({\bf r})|^2d^2r=N$,
the normalization constant is $A=N^{1/2}/a$.  
The number of excitons may  be estimated from $N=(2\pi Rd)\epsilon_F D_e$, 
where $d\simeq 1\sim 10\mu  {\rm m}$ is the width of the ring, and $D_e$ and $\epsilon_F$
are the electron density of states
and the Fermi energy respectively.
It is useful to write the GPE  in a dimensionless form:
\begin{eqnarray}
&&-\frac{1}{2}\nabla^2\Psi({\bf x})+\frac{1}{2}\eta_0(x^2-x_0^2)^2\Psi({\bf x})
+V_D({\bf x})\Psi({\bf x}) \nonumber \\
&&+\eta_1|\Psi({\bf x})|^2\Psi({\bf x})-\eta_2|\Psi({\bf x})|^4\Psi({\bf x})
+\eta_3|\Psi({\bf x})|^6\Psi({\bf x}) \nonumber \\
&&=\beta \Psi({\bf x}) .
\end{eqnarray}
Here $\eta_0=\omega/\Omega$,   $x_0=R/a$, and $\beta=\mu_B/\hbar\Omega$.
The coefficients of the nonlinear terms are
$\eta_1=\frac{A^2}{2D}\frac{1}{\hbar\Omega}=
\pi(\frac{\epsilon_F}{\hbar\Omega})(\frac{dR}{a^2})\frac{m_e}{m^*}$,
$\eta_2=
\frac{3\pi^2}{2}\frac{\epsilon_F}{\hbar\Omega}(\frac{\epsilon_F}{|\mu|})
(\frac{dR}{a^2})^2(\frac{m_e}{m^*})^2$, and 
$\eta_3=\frac{2\pi^3}{3}\frac{\epsilon_F}{\hbar\Omega}
(\frac{\epsilon_F}{|\mu|})^2(\frac{dR}{a^2})^3(\frac{m_e}{m^*})^3$. 
We use  $a=10\mu {\rm m}$, which gives 
$\hbar \Omega \simeq 10^{-6}{\rm meV}$.
For   exciton densities between $10^{9}\sim 10^{10}{\rm cm}^{-2}$ 
we estimate  $\eta_1\simeq 10^4\sim 10^6$.  These large values will have important consequenses (see below).
Since $|\mu|$ is comparable to the indirect exciton binding energy of few meV
we get  $\epsilon_F/|\mu|< 0.01$.  From these considerations, we conclude that
the nonlinear terms, except  the third order term, may be neglected. Note that the 
parameter $\beta$, {\it i.\ e.,} the binding energy $\epsilon_0$ is determined from the
solution of the GPE.

\begin{figure}
\includegraphics[width=0.5\textwidth]{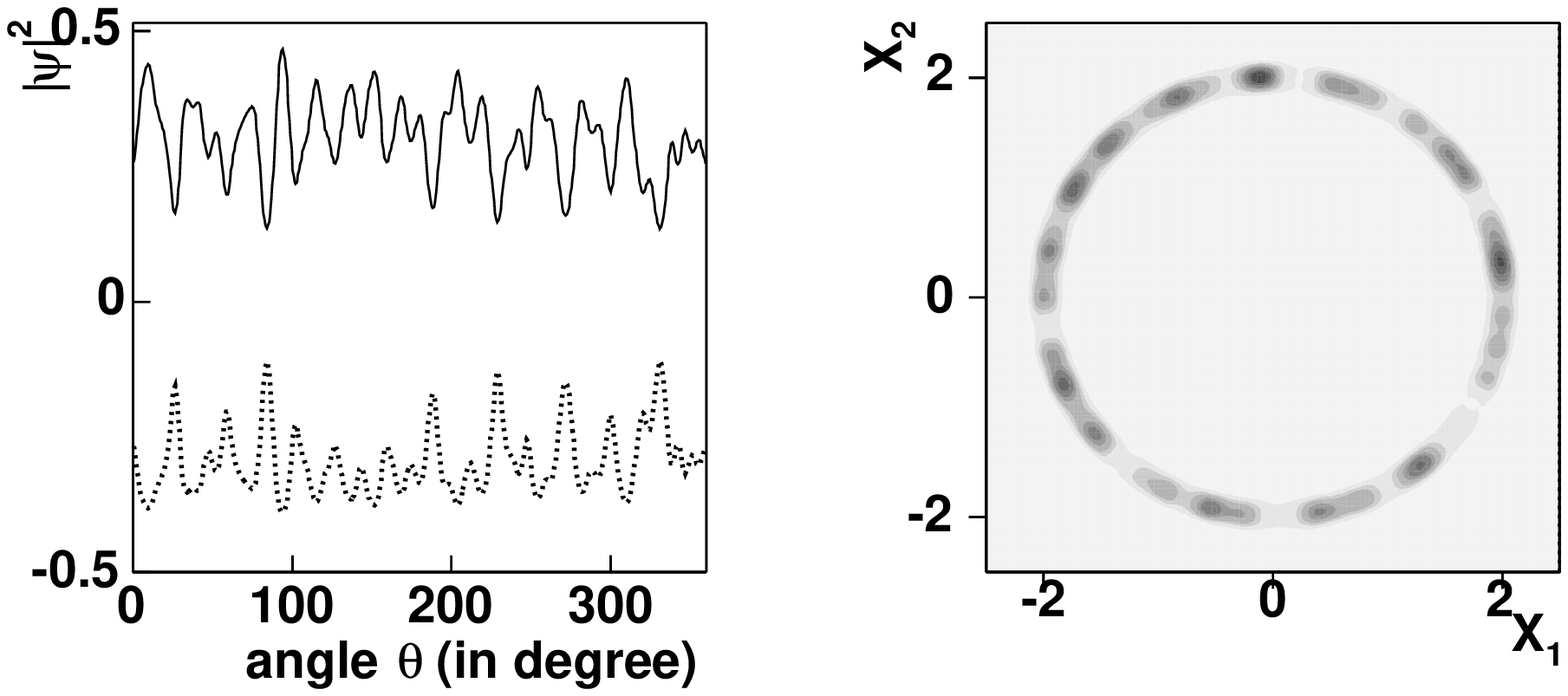}
\includegraphics[width=0.5\textwidth]{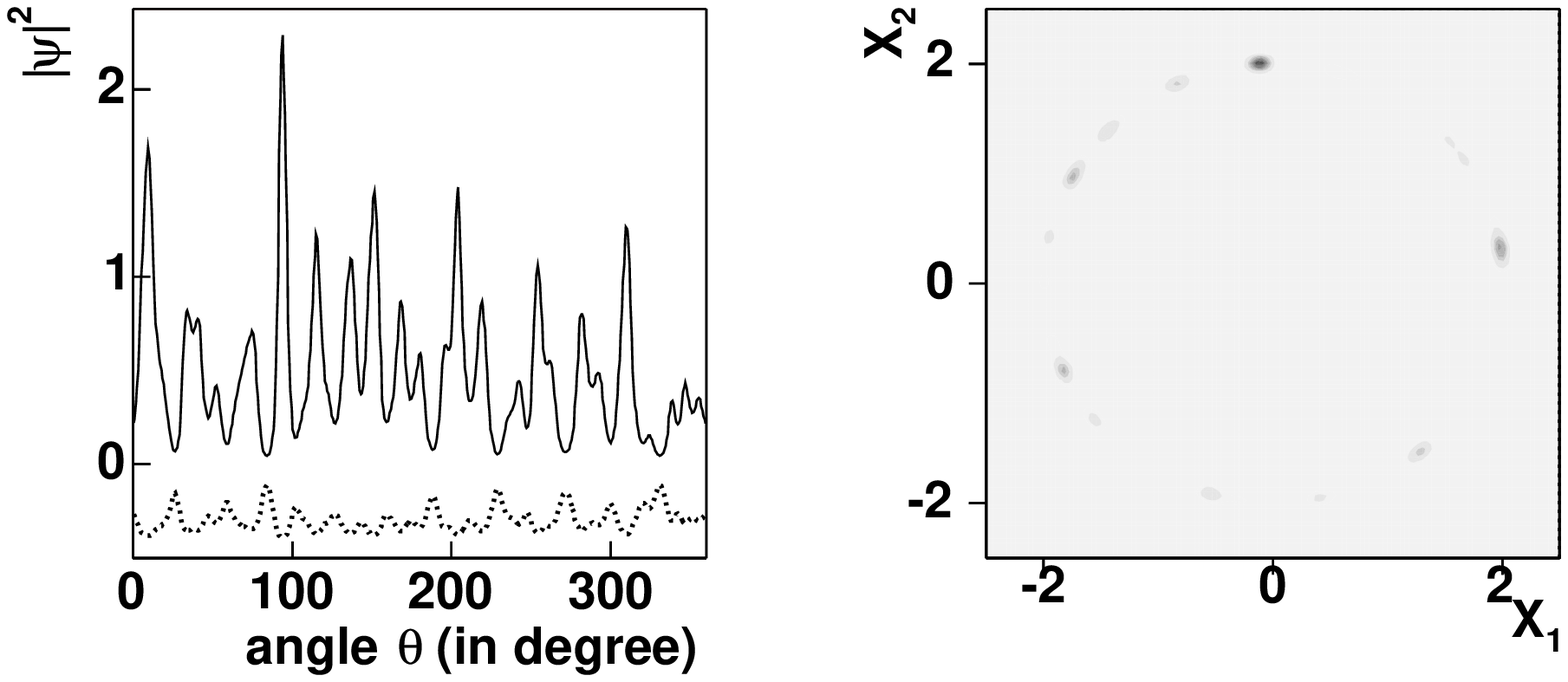}
\caption{Upper panels: Left figure displays the
variation of $|\Psi({\bf x}) |^2$ along the circumference of the ring.  
The dashed line shows the random potential $V_D({\bf x})/50$
measured in units of $\hbar\Omega$. The values of $V_D$ are shifted
by an arbitrary amount for clarity. The third-order nonlinear coefficient is $\eta_1=10^{5}$.  
Right figure shows a two-dimensional contour plot of $|\Psi({\bf x})|^2$.
Darker regions represent larger values of 
$|\Psi({\bf x}) |^2$. The two-dimensional position vector is $\vec{x}=(x_1,x_2)$ measured in units of 
$a=10\mu {\rm m}$.
Lower panels: Same as in the upper panel but with  $\eta_1=0$. Note
that  the same random  potential is used
as in the upper panels. Only the y-axis scale is different.
As $\eta_1$ increases fluctuations of peak values of  $|\Psi({\bf x}) |^2$ are suppressed,
but the width of the ring increases.}
\label{geometry1}
\end{figure}

We now solve the dimensionless GPE for the ground
state using the imaginary time evolution method.
The two-dimensional space is discretized using
$512\times512$ grid points on a square of area $10a\times10a$.
The random potential is simulated by  $250$ Gaussian potentials 
placed randomly on the ring:  $V_D(\vec{r})=\sum_i Ve^{-(\vec{r}-\vec{R}_i)^2/w^2}$, where $\vec{R}_i$
are random position vectors on the ring with radius $R$ and  $w=\frac{a}{\sqrt{50}}\sim 1 \mu {\rm m}$.
The strength
of the random potential is taken to a fraction of the strength of the confinement potential
$V=0.25\hbar\omega$.
The correlation
function of the random potential is $ \overline{  V_D(\vec{x}) V_D(\vec{x'}) }
\propto \frac{1} {2\pi w^2} 
e^{ -|\vec{x}-\vec{x'}|^2/2w^2   }$.  
We find numerically that the average distance between the peaks in  $|\Psi({\bf x})|^2$
is roughly the correlation
length of the random potential $w$.  
The upper left panel of Fig.~1
displays how $|\Psi({\bf x})|^2$ varies on the circumference of ring for  $\eta_1=10^{5}$
and $\hbar\omega=2\times10^5\hbar\Omega$.
A two-dimensional contour plot of $|\Psi({\bf x})|^2$ displays a necklace-like structure as shown
in the upper right  panel of Fig.1.
Positions of modulation maxima are nearly periodic while the values of modulation maxima
are non-periodic.   The angular distances between 
peak positions are given roughly as $12-20^\circ$, which shows
a deviation from the mean period roughly about 20 percent.
These results are in qualitative agreement with the experimental data (see Fig.3c of
Butov et al \cite{butov2}).   
The width of the ring increases with increasing coefficient of
the non-linear 
repulsive term $\eta_1$.
In the lower panels of Fig.~1, 
the solution to GPE is displayed when the repulsive third order term is absent while the disorder 
potential is present.  We observe that the modulation maxima fluctuate much more than those 
of $\eta=10^5$, and that  periodic modulation is absent in a two-dimensional contour plot 
of $|\Psi({\bf x})|^2$.
Comparing $|\Psi({\bf x})|^2$ with the shape of the random potential
in the left panels of Fig.~1, we can conclude that,
as the repulsive third order term increases, 
the peak
values of $|\Psi({\bf x})|^2$ in the local minima of the potential fluctuate less from each other.
We have also solved the GPE when  the repulsive third order term is present while the disorder 
potential is absent.  In this case, $|\Psi({\bf x})|^2$ is uniform along the center of the ring, which
implies that 
the uniform ring state represents 
a stable BEC of excitons in the absence of disorder.

We have derived a nonlinear Gross-Pitaevkii equation (GPE) of exciton BEC  in the presence 
of  small random variation of ring width, and have shown that the ground state 
displays a necklace-like structure.  Our results demonstrate that  the interplay between random potential
and the nonlinear repulsive term of  the 
GPE plays an important  role in understanding necklace-like structures.
Since the observed fragmented structure appears suddenly at low temperature,
a collective mechanism of quantum origin may be involved, such as the BEC.
Since we assumed a  BEC phase of excitons in our derivation  
our results may be especially relevant in the low temperature regime of the phase diagram of the ordered
state proposed
by Butov et al \cite{butov3}.
We believe more experiments at lower temperature
are needed that would 
elucidate the interplay between the non-linear repulsive term and 
random potential in the formation of  the fragmentation. 

SREY thanks D. Chemla and L.V. Butov for useful conversations.
This work is supported by Korea Research Foundation   
grant KRF-2003-015-C00223, and  by grant No.(R01-1999-00018) from the interdisciplinary
research program of the KOSEF, and by KOSEF R01-2001-00016.


\end{document}